\documentclass[%
 reprint,
 amsmath,amssymb,
 aps,superscriptaddress, prl
]{revtex4-1}

\usepackage{graphicx}% Include figure files
\usepackage{dcolumn}% Align table columns on decimal point
\usepackage{bm}% bold math
\usepackage{siunitx}
\usepackage{color}

\usepackage[symbol]{footmisc}

\usepackage[final]{pdfpages}
\usepackage{etoolbox} 

\makeatletter
\patchcmd{\@outputpage@head}{\@ifx{\LS@rot\@undefined}{}{\LS@rot}}{}{}{}
\makeatother

\usepackage{afterpage}

\begin{document}

\title{Topological Thermoelectricity in Metals}

\author{Sobhit Singh}
\affiliation{Department of Physics and Astronomy, West Virginia University, Morgantown, WV-26505-6315, USA}

\author{QuanSheng Wu}
\thanks{S. Singh and Q. Wu equally contributed to this work.}
\affiliation{Institute of Theoretical Physics, Ecole Polytechnique F\'ed\'erale de Lausanne (EPFL), CH-1015 Lausanne, Switzerland}
\affiliation{Theoretical Physics and Station Q Zurich,  ETH Zurich, CH-8093 Zurich, Switzerland}

\author{Changming Yue}
\affiliation{Beijing National Laboratory for Condensed Matter Physics, and Institute of Physics, Chinese Academy of Science, Beijing 100190, China}

\author{Aldo H. Romero}
\affiliation{Department of Physics and Astronomy, West Virginia University, Morgantown, WV-26505-6315, USA}

\author{Alexey A. Soluyanov}
\affiliation{Theoretical Physics and Station Q Zurich,  ETH Zurich, CH-8093 Zurich, Switzerland}
\affiliation{Physik-Institut, Universit\"at Z\"urich, Winterthurerstrasse 190, CH-8057 Zurich, Switzerland}
\affiliation{Department of Physics, St. Petersburg State University,  St. Petersburg, 199034, Russia}


%\affiliation{
%\affiliation{
%\affiliation{%


\date{\today}% It is always \today, today,

\begin{abstract}
Recently published discoveries of acoustic and optical mode inversion in the phonon spectrum of certain metals became the first realistic example of non-interacting topological bosonic excitations in existing materials. However, the observable physical and technological use of such topological phonon phases remained unclear. In this work we provide a strong theoretical and numerical evidence that for a class of metallic compounds (known as triple point topological metals), the points in the phonon spectrum, at which three (two optical and one acoustic) phonon modes (bands) cross, represent a well-defined topological material phase, in which the hosting metals have very strong thermoelectric response. The triple point {\it bosonic} collective excitations appearing due to these topological phonon band-crossing points significantly suppress the lattice thermal conductivity, making such metals {\it phonon-glass} like. At the same time, the topological triple-point  and Weyl {\it fermionic} quasiparticle excitations present in these metals yield good electrical transport ({\it electron-crystal}) and cause a local enhancement in the electronic density of states near the Fermi level, which considerably improves the thermopower. This combination of ``{\it phonon-glass}" and ``{\it electron-crystal}" is the key for high thermoelectric performance in metals. We call these materials topological thermoelectric metals and propose several newly predicted compounds for this phase (TaSb and TaBi). We hope that this work will lead researchers in physics and materials science to the detailed study  of topological phonon phases in electronic materials,  and the possibility of these phases to introduce novel and more efficient use of thermoelectric materials in many everyday technological applications. 
\end{abstract}

\maketitle

By now, an almost complete classification of possible topological \textit{fermionic} excitations in non-interacting systems has been constructed~\cite{Bradly_science2016, Bradly_nature2017}, and material examples hosting many of these classified phases are ubiquitous. This classification includes the so-called triple-point metals~\cite{ZhuTPM2016, HWengPRB2016, Winkler_PRL2016, ZaheerPRB13,BQLi_Nature2017}, in which a symmetry-protected crossing of three bands occur close to the Fermi level, resulting in topological fermionic excitations.  Here we claim that electronic crystalline compounds also have topologically protected {\it bosonic} excitations that result in non-standard thermoelectric properties of the hosting materials.     

Topological bosonic modes were proposed and realized in classical mechanical systems~\cite{Roman_science2015, SusstrunkPNAS2016, Huber2016}, where oscillations of pendulums are although purely classical, but realize the same physics, as some quantum electronic systems do~\cite{KaneNature2014, Stenull_PRL2016}. Another realization of topological bosonic states came from experiments on photonic crystals~\cite{Khanikaev_Natu2013, Ling_NaturePho2014, KaiSun_Nature2012, JMei_PRB2012, Raghu_PRB2008, LingLu_Nature2013, Rechtsman_PRL2013, Wen_Scireports2015}, providing a realization of Dirac, Weyl, and nodal line phases of bosons in a \textit{bosonic} system. Several very recent works described the realization of topological bosonic modes in \textit{electronic} systems~\cite{zhangPRL2017double, Li-PRB18, EsmannPRB2018}. For example, the appearance of a bosonic triple-point analogue of triple point fermion in the phonon spectrum of several existing compounds is discussed in Ref.~\cite{Li-PRB18}. Phononic Weyl points were also reported in these metallic materials. Nonetheless, unlike the observable effects of topological fermionic excitations, no observable physical effect of these topological bosonic excitations was predicted. 

The topological character of bosonic (phononic) Weyl points can be easily proven following the direct analogy with fermionic Weyl points. In fact, not only the methodology of computing chiralities of electronic Weyl points can be used for their phonon analogues, but even the same, already developed, software for fermionic topological excitations~\cite{Z2Pack, wu2017wanniertools} reveals the topology in this case. However the topological origin of the triple point in the phonon spectrum is more complicated to capture. 

Here we provide the proof of the topological nature of a triple-point phonon. We \textit{add several novel metals} to the compounds described in Ref.~\cite{Li-PRB18}. Most importantly, we provide the topology-mediated observable physical effect, proving that the presence of triple-point phonons makes these topological metals the most efficient thermoelectric metals of those known to date. We show in details that the enhanced thermoelectric response is driven by the topological band-crossings present in the phonon spectrum. 

In this work we investigated the electronic and vibrational properties of nine compounds from Ta$X$ and Nb$X$ ($X$ = N, P, As, Sb, Bi) families, considering all of them to have the tungsten carbide (WC) type crystal structure (space group 187). The lattice parameters, electronic band structures and phonon mode structures of all these compounds are given in supplemental information (SI)~\cite{SI_file}. Our calculations reveal that TaP, TaAs, TaSb, TaBi, NbP, NbSb and NbBi compounds exhibit triple-point and Weyl \textit{fermionic} excitations, whereas, TaN and NbN compounds only have triple-points in their electronic spectrum, as shown in SI~\cite{SI_file}. 

The \textit{phonon} structure of all these compounds is similar, except for TaSb and TaBi compounds, in which an optical and acoustic mode inversion takes place in the phonon spectrum. In the electronic spectrum a similar band inversion between occupied and unoccupied states is a signature of the non-trivial topology. The same is true for crossings of optical and acoustic phonon modes, as we prove below. We will show that the presence of such topological phonon mode crossings (i.e. triple-point phonons) significantly suppresses the lattice thermal conductivity, and thereby enhances the thermoelectric performance of these special topological metals. 

\begin{figure}[htb]
 \centering
 \includegraphics[width= 9cm]{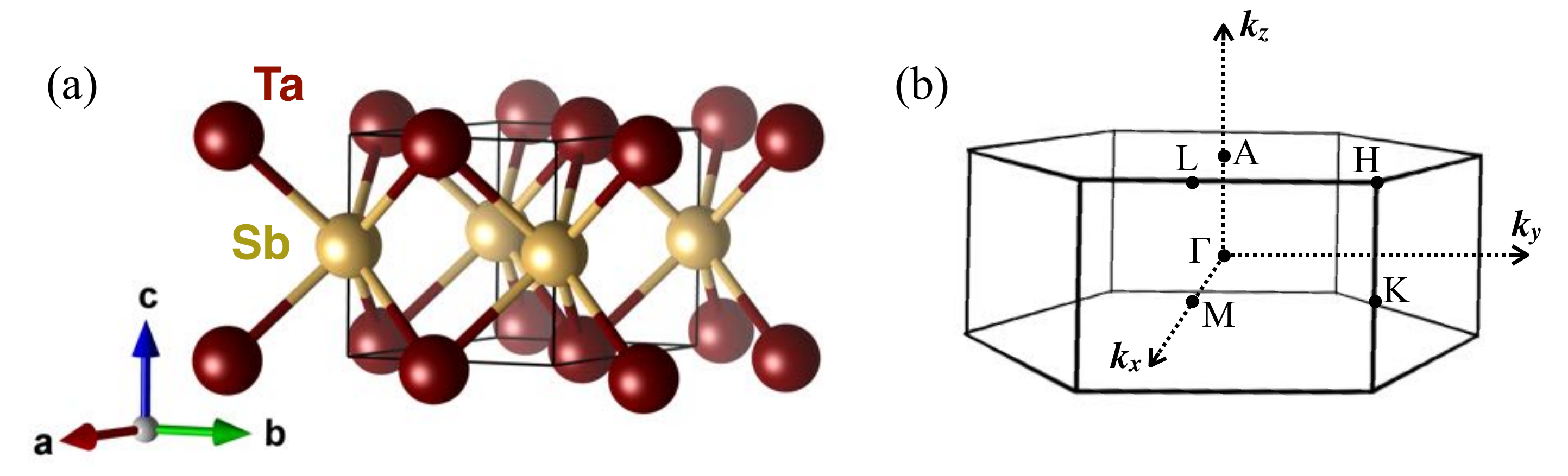}
 \caption{(Color online) (a) Crystal structure of TaSb compound in WC-type $P\bar{6}m2$ space group. Ta atoms (maroon) occupy the (2/3, 0.0, 2/3) site, and Sb atoms (golden) occupy the (0.0, 1/2, 0.0) site. The unit cell contains one Ta and a one Sb atoms. The analysis of the band structure topology (type-A triple points and several Weyl points found in the spectrum) is presented in the SI along with the full exposition of the phonon spectrum. (b) Brillouin zone of TaSb.}
 \label{fig:Atom_structure}
 \end{figure}

%In the following, we choose the novel compound TaSb as a prototype for the detailed illustration of the topological features in vibrational properties. The lowest energy crystal structure of TaSb is shown in Fig.~\ref{fig:Atom_structure}. Our calculations (see SI) indicate that this structure is energetically, vibrationally, mechanically and thermodynamically stable at 0 K and at room temperature. 
In the following, we choose the novel compound TaSb as a prototype for the detailed illustration of the topological features in vibrational properties. The lowest energy crystal structure of TaSb is shown in Fig.~\ref{fig:Atom_structure}. Our calculations (see~\cite{SI_file}) indicate that this structure is stable at 0 K and at room temperature. 
 
All the considered compounds have two atoms per primitive unit cell, hence having six phonon bands (three acoustic and three optical) in the spectrum. It is shown in the SI that the phonon triple point appears in those compounds, where the masses of atoms forming the material, satisfy the following condition for phonon mode inversion: The frequency gap $\Delta$ between the acoustic and optical phonon bands at the A-point of Brillouin zone (BZ) has to become negative for these bands to invert 
\begin{equation}
   \Delta \propto \sqrt{\beta_{\parallel}} - \sqrt{\beta_{\perp}\frac{m}{M}}.
\end{equation}
Here, $\beta_{\parallel}$ and $\beta_{\perp}$ are the in-plane and out-of-plane second-order interatomic force constants between the atoms of masses $m$ and $M$ ($m < M$), respectively. 

This relationship is verified by our calculations: $\Delta$ decreases systematically with the increase of $m/M$ ratio. For TaN ($m/M$ = 0.0774), TaP ($m/M$ = 0.1711), and TaAs ($m/M$ = 0.414) compounds the obtained $\Delta$ values are $200.5$, $115.1$, and $14.4$ cm$^{-1}$, respectively, but for TaSb ($m/M$ = 0.6731) and TaBi ($m/M$ = 0.8658) compounds $\Delta$ becomes negative (-16.7 and -27.8 cm$^{-1}$ correspondingly) indicating the inversion of phonon modes in the BZ. In addition to increasing $m/M$ ratio, $\beta_{\perp} > \beta_{\parallel}$ condition is essential to observe the phonon mode-inversion in the BZ. Results of Li et al.~\cite{Li-PRB18} further corroborate the aforementioned relationship (see~\cite{SI_file}). 

\begin{figure*}[htb]
 \centering
 \includegraphics[width=18cm]{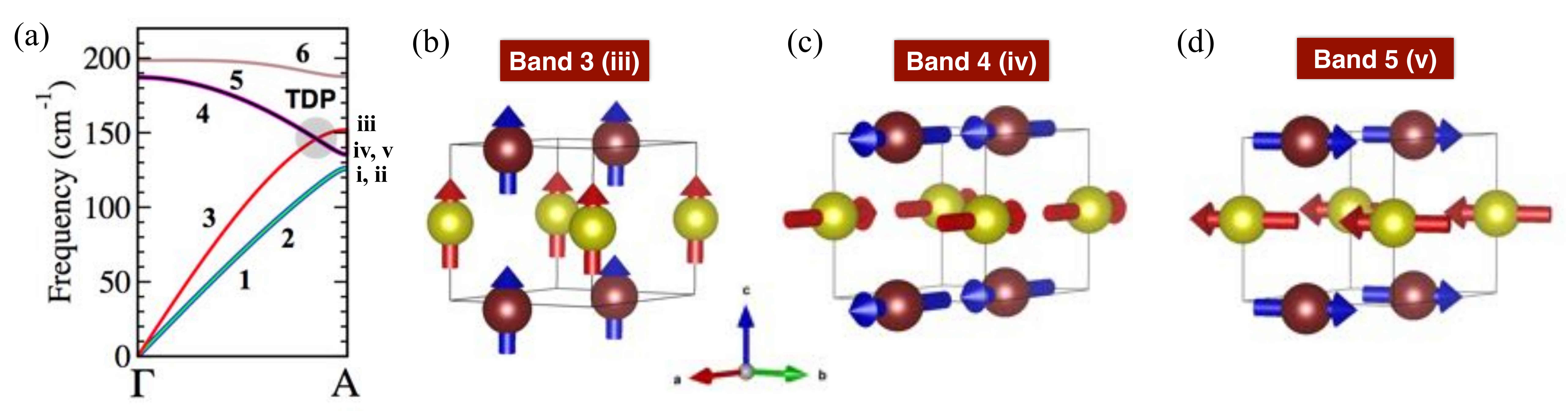}
 \caption{(Color online)(a) Part of the phonon spectrum of TaSb. The triply-degenerate point (TDP), at which phonon bands 3(iii), 4(iv) and 5(v) meet is the topologically protected triple-point of phonons (another such point is located in the other direction in the BZ on the same $C_{3v}$-symmetric line). Panels (b, c, d) explicitly illustrate the atomic vibrations corresponding to the overlapping phonon modes 3(iii), 4(iv) and 5(v) correspondingly. The enumeration of phonon bands is made with numbers (1..6) and (i..vi) ``before" and ``after" the crossing point, to stress the change in phonon band ordering by using different number styles for ordinary and inverted phonon band orderings.}
 \label{fig:phonons}
 \end{figure*}

We thus conclude that while the electronic properties of TaSb in $P\bar{6}m2$ group are similar to those of other triple-point-metals of Ta$X$ and Nb$X$ family ($X$ = N, P, As), the phonon spectrum of TaSb and TaBi compounds makes them distinct from the other triple-point-metals in the considered material set. In particular, we observe that one acoustic band (no. 3) and two degenerate optical bands (no. 4 and 5) in TaSb and TaBi compounds intersect each other along the $\Gamma$-A path forming two triply-degenerate phonon points (TDP) as shown in Fig.~\ref{fig:phonons}(a) for TaSb. The atomic vibrations corresponding to these phonon bands are illustrated in Fig.~\ref{fig:phonons}(b-d). The TDPs are located at frequency $\approx$145 cm$^{-1}$ and at ${\bf q}=(0, 0, \pm 0.428)$.  The phonon mode inversion can be noticed above the TDP (in $k_z$), where two degenerate optical phonon bands (no. iv and v) unusually appear at lower frequencies than the acoustic phonon band (no. iii). The phonon mode inversion, at the high-symmetry point A(0, 0, $\pm\frac{\pi}{c}$),  is indicative of the non-trivial topological nature of the vibrational properties in TaSb compound. Despite the isoelectronic similarity of all the considered compounds, such phonon band-inversion along $\Gamma$-A path only in TaSb and TaBi, is due to the fact that atoms of relatively similar masses constitute these compounds. 

 Topological classification of non-interacting fermionic excitations is built from the symmetry analysis of tight-binding models. Analogously, topological classification of phonons (bosonic excitations) can be built from the analysis of dynamical matrices (a good, although not fully complete in terms of various symmetry constraints, exposition of this approach is presented for mechanical vibrations in Ref.~\cite{SusstrunkPNAS2016}). 
 
Starting from this point of view, let us now prove that the TDP indeed has a topological origin, for which a topological invariant can be defined. To capture the presence of fermionic triple points in the band structure, two triple points are enclosed into a surface in the BZ. Computation of the Wilson loop of lower-lying bands on this surface gives a robust topological signature of triple fermionic points in metals~\cite{ZhuTPM2016}. Although TDPs also appear in pairs in the BZ, unlike the case of Weyl points, a direct analogy to the fermionic approach to obtain the topological invariant does not work for TDPs. 

However, we can still find the path to define a topological invariant, if we keep in mind that type-A triple points in electronic band structure are connected by a single nodal line, which is topologically trivial, having the $0$ value of the Berry phase ($\phi_B$) of occupied bands on any circular path enclosing it. Nonetheless, the hidden topology is revealed by applying the Zeeman field~\cite{ZhuTPM2016}, which splits the two triple points into four Weyl points. 

\begin{figure*}[htb]
 \centering
 \includegraphics[width=18cm]{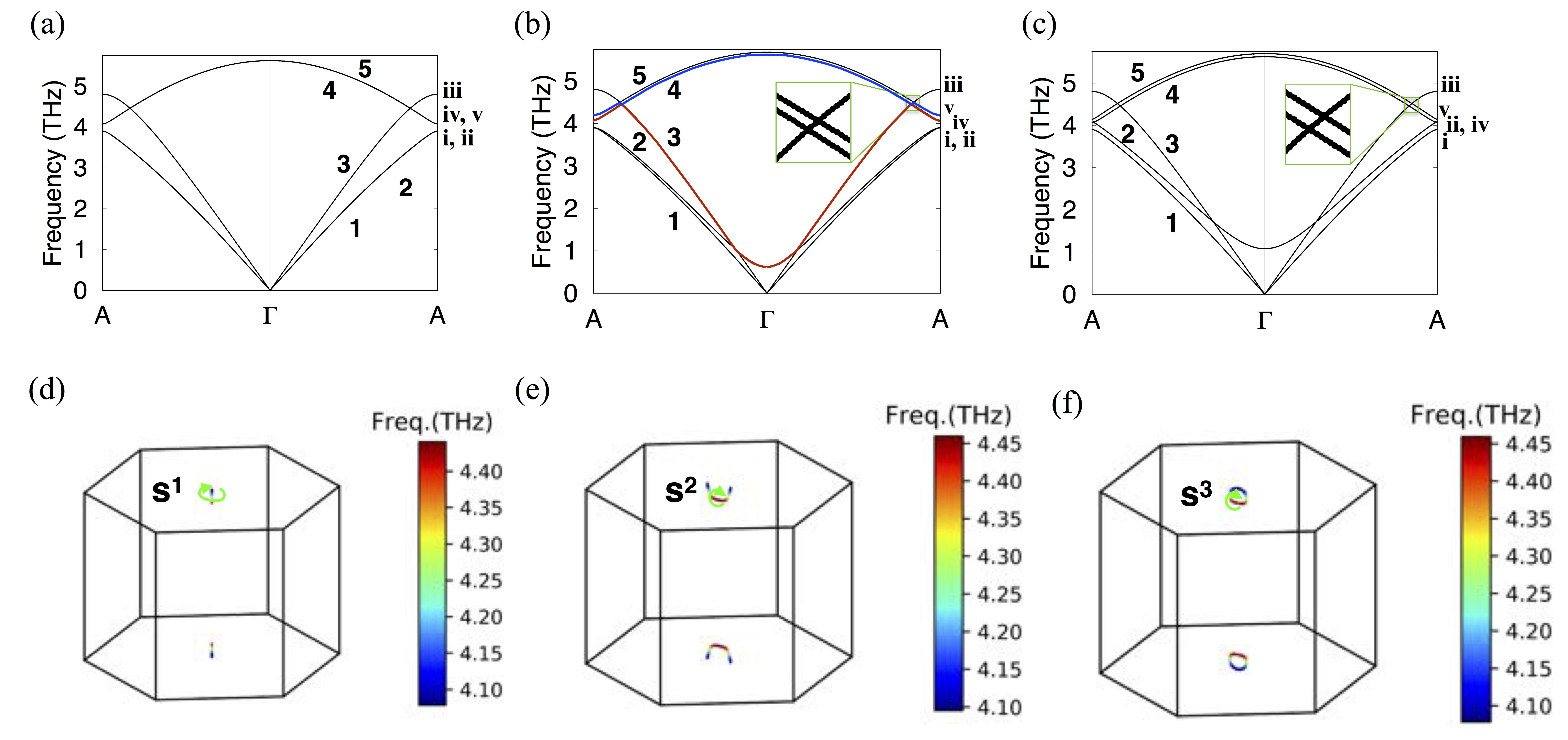}
 \caption{(Color online) (a) The phonon spectrum of TaSb along the $C_{3v}$-symmetric line of the BZ: Doubly degenerate phonon bands (4,5) and (iv,v) can be seen. (b) The phonon spectrum of TaSb with the $xx$ and $yy$ force constants in the dynamical matrix made unequal for Ta. The degeneracy of (4,5) and (iv, v) is lifted. (c) The phonon spectrum of TaSb with the $xx$ and $yy$ force constants in the dynamical matrix made unequal for Sb. The degeneracy of (4,5) and (iv, v) is lifted. (d) The non-topological \textit{open nodal line} formed in the BZ by bands iv and v of panel (a). The loop $S^1$ encircles this line, but  $\phi_B(S^1)=0$. (e) The \textit{closed nodal loop} formed by bands iv and v in the BZ for the case of panel (b) away from the $\Gamma$-A line. The contour $S^2$ links with this loop and $\phi_B(S^2)=\pi$. (f) The \textit{closed nodal loop} formed by bands iv and v in the BZ for the case of panel (c). For a contour $S^3$ linked with this loop $\phi_B(S^3)=\pi$.
 } 
 \label{fig:topology}
 \end{figure*}

We apply a similar trick here to capture the topology of TDPs. Instead of the Hamiltonian in the electronic case, we deal with the dynamical matrix, the eigenvalues of which, similar to the type-A triple point electronic Hamiltonian case, connect the two TDPs by a degenerate line (this line in Fig.~\ref{fig:phonons}(a) is formed by the degenerate phonon bands iv and v).  The degeneracy between phonon bands 4 and 5, as well as between bands iv and v, occurs due to the presence of $C_{3v}$ symmetry in the considered crystal lattice, because the $xx$ and $yy$ force constants transform like $x$ and $y$ coordinates. The latter transform in the same way as $p_x$ and $p_y$ orbitals that in the $C_{3v}$ group form a 2-dimensional irreducible representation $E$, thus enforcing the degeneracy seen in Fig.~\ref{fig:topology}(a).  

The degeneracy line (not a loop, but an open line) in the BZ of bands iv and v is illustrated in Fig.~\ref{fig:topology}(d).
 Analogously to the nodal line in type-A electronic triple-points, this line of degenerate modes iv and v (on the right side of the crossing point in Fig~\ref{fig:phonons}(a)) carries $0$ Berry phase ($\phi_B$) computed for the lower-lying phonon bands around the path $S^1$ of Fig.~\ref{fig:topology}(d). To reveal the topology of the TDP we use the trick similar to Zeeman splitting mentioned above. To do that, we modify the entries in the dynamical matrix in a way that corresponds to changing the semi-classically understood ``spring constants" connecting atoms along the $x$ and in the $y$ directions (that can be related to the corresponding interatomic bond strength). This modification splits the degeneracy of the phonon nodal line, as illustrated in Figs.~\ref{fig:topology}(b, c).
 
 It turns out that such a splitting gaps the phonon bands 4(iv) and 5(v) along the ($\Gamma$-A) direction of the BZ. However, away from this line we still see the nodal \textit{loops} formed by the inverted bands iv and v. Computation of $\phi_B$ of the lower-lying phonon bands along the path $S^2$ linked with this degeneracy loop gives  $\phi_B=\pi$, thus proving the topological nature of the TDPs, just like the splitting of electronic triple points into four Weyl points by the Zeeman field proves the topological origin of type-A fermionic triple-points.

Despite this proof of the topological origin of TDPs, the more important question is the observable physical consequences, associated with these topological phonon modes.  Although our calculations of the surface phonon spectrum of TaSb reveal topological surface modes, they are not really observable, being heavily intermixed with the bulk modes~\cite{SI_file}. The work of Ref.~\cite{Li-PRB18} demonstrated evidence of similar non-trivial surface modes (open Fermi arcs) in TiS and HfTe compounds that also host TDPs in their phonon spectra. However, similar to the case of TaSb, the phonon surface modes in these compounds are also buried in the projection of bulk phonon modes. For this reason, we explore another possible signature of non-trivial topological excitations, that is, transport properties. Of these properties, the one closely related to phonons (although with a large contribution from electrons as well), is the thermal transport~\cite{NielsenPRB1974}. For this reason we study the effect of the phonon TDPs on the thermoelectric properties of hosting metals, and find them to be among the most \textit{efficient metallic thermoelectrics}.  

The performance of thermoelectric materials at temperature $T$ is usually determined by the figure of merit, $zT$ = $\frac{S^{2}\sigma}{\kappa_{el} + \kappa_{ph}}T$, where quantities $S$ and $\sigma$ are Seebeck coefficient and electrical conductivity, whereas, $\kappa_{el}$ and $\kappa_{ph}$ are the electronic and phononic (lattice) contributions to the thermal conductivity of material. A combination of large $S$ and large $\sigma$ together with a small $\kappa$ ($=\kappa_{el} + \kappa_{ph}$) is required to achieve large $zT$. In other words, a good thermoelectric is a good conductor of charge carriers (i.e. ``electron-crystal") and a bad conductor of phonons (i.e. ``phonon-glass"). Thus, in case of metals hosting coexisting topological electron and phonon excitations, we expect a significant benefit towards the enhancement of $zT$. 

\begin{figure}[htb]
 \centering
 \includegraphics[width=9cm]{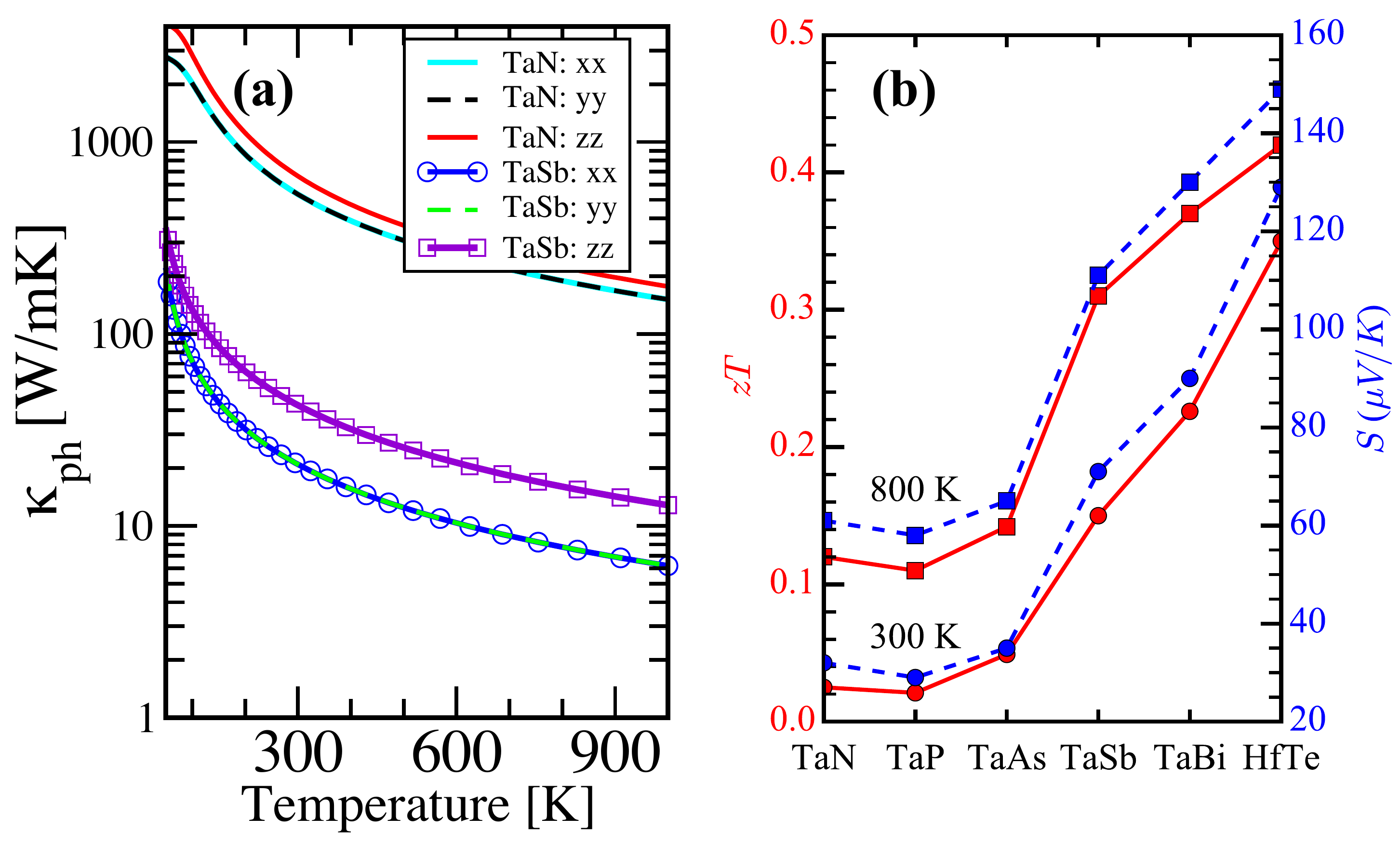}
 \caption{(Color online) (a) Variation of different components of lattice thermal conductivity tensor ($\kappa_{ph}$) against temperature T for TaN and TaSb along $x$, $y$ and $z$ directions of crystal. (b) Comparison of maximum $zT$ (red) and Seebeck coefficients (blue) obtained for electron doping at 300~K (circles) and 800~K (squares). } 
 \label{fig:thermal}
 \end{figure}

First, the presence of topologically protected phonon band-crossings (TDPs) considerably suppress the lattice thermal conductivity due to the increased phonon scattering centers. The $\kappa_{ph}$ (and the phonon mean free path) in TaSb is almost two orders in magnitude smaller compared to that of TaN which is a metal without TDPs (see Fig.~\ref{fig:thermal} and more details in SI). According to the Wiedemann-Franz law, $\kappa_{el} = L_{0} T \sigma$, where $L_{0}$ is the Lorenz number and $\tau$ is the electron-phonon relaxation time~\cite{Snyder2008}. Since $\sigma/\tau \sim 10^{20} ~1/{\si{\ohm}}ms$ at low-doping concentrations for all the studied metals, the $\kappa_{el}$ (which mainly governs $\kappa$ at high $T$) is also in the same order of magnitude for all TaX compounds. Therefore, reduction in $\kappa_{ph}$ (which dominates $\kappa$ at low $T$) causes a net decrease in $\kappa$ of TaSb and TaBi~\cite{SI_file}. 

Second, there is an increase in the Seebeck coefficient ($S$) of TaSb and TaBi. In particular, we notice a local enhancement in the electronic density of states (DOS) near the Fermi level as we go from TaN to TaBi. This occurs due to the flattening of electronic bands at the Fermi level in TaSb and TaBi compounds~\cite{SI_file}. According to the Mahan-Sofo theory~\cite{Mahan7436}, such a situation yields an increase in $S$, which in turn improves $zT$. In general, the sharper is the local increase in DOS at the Fermi level, the larger is the enhancement in $S$. Decrease in $\kappa_{ph}$ and an increase in $S$ combine to improve the overall thermoelectric efficiency $zT$.

In Fig.~\ref{fig:thermal}, we compare the $zT$ of all the TaX family compounds considered in this work, as well as one of the previously predicted TDP compounds HfTe~\cite{Li-PRB18}. Although the illustrated $zT$ of these compounds is smaller than that of the best known \textit{narrow bandgap semiconductor thermoelectrics} (having $zT\approx 1$), it is among the best known \textit{thermoelectric metals} and performs \textit{much better} compared to non-topological metals, for which $zT$ ranges from $0.0001$ -- $0.001$~\cite{TerasakiPRB1997, TakahataPRB2000, OkudaPRB2001}. Notably, we find that HfTe, a triple-point metal hosting TDPs predicted in Ref.~\cite{Li-PRB18}, exhibits considerably large $zT$ (maximum $zT$ = 0.35 at 300~K and 0.42 at 800~K for electron doping case) compared to that of TaSb and TaBi compounds due to its relatively large Seebeck coefficient (maximum $S$ = 129 $\mu V/K$ at 300~K and 149 $\mu V/K$ at 800~K for electron doping). The results that we obtain here for TaSb, TaBi, and HfTe triple-point metals, put these materials in the list of the best thermoelectric metals known to date, along with a lower symmetry (space group $\#109$) TaAs~\cite{PENG2016225} and recently reported HgTe~\cite{ThermHgTe}. 

In summary, our work provides the first illustration of the use of topological phases of phonons in existing electronic compounds and predicts novel materials TaSb and TaBi, which combine electronic and bosonic (phononic) topological phases. The associated topological excitations - fermions and phonons - give rise to enhanced thermopower and thermoelectric response in considered metals. The topological phonon modes considerably suppress the lattice thermal conductivity without disrupting the electronic transport in the bulk crystal, whereas the topological electronic bands near the Fermi level give rise to a local increase in the DOS at the Fermi level causing an increment in the thermopower. Having illustrated the approach to prove the topological phase of phonon spectrum and the role of non-trivial phonon modes on the thermal transport properties, we hope this work will motivate further studies of the effects of topologically non-trivial phonons on the material properties, including thin films  and metal/semiconductor heterostructures, where the effects of topology and phonon scattering can be more pronounced. \\

%%%%%%%%%%%%%%%%%%%%%%%%%%%%%%%%%%%%%%%%%%%%%%%%%%%%%%%%%%%%%%%%%%%%%

\bibliography{arxivbib.bib}

\noindent\textbf{\\ \\ Acknowledgements\\} 
This work used the Extreme Science and Engineering Discovery Environment (XSEDE), which is supported by NSF grant OCI-1053575. Additionally, we acknowledge support from Texas Advances Computer Center (TACC), Bridges supercomputer at Pittsburgh Supercomputer Center and Super Computing Systems (Spruce and Mountaineer) at West Virginia University. A.H.R. and S.S. acknowledge the support from National Science Foundation (NSF) DMREF-NSF 1434897, OAC-1740111 and DOE DE-SC0016176 projects. S.S. acknowledges the support from the Dr. Mohindar S. Seehra Research Award and the Robert T. Bruhn Research Award at West Virginia University. Q.S.W and A.A.S. acknowledge the support of Microsoft Research, Swiss National Science Foundation (SNSF) NCCR QSIT and NCCR MARVEL grants. A.A.S. acknowledges the support of the SNSF Professorship grant. \\

\noindent\textbf{ORCID id\\}
Sobhit Singh: 0000-0002-5292-4235 \\
QuanSheng Wu: 0000-0002-9154-4489 \\
Changming Yue:  0000-0002-2651-5139 \\
Aldo H. Romero: 0000-0001-5968-0571 \\
Alexey A. Soluyanov: 0000-0003-3539-1024 \\
 
\noindent\textbf{Corresponding authors: \\}
Sobhit Singh: smsingh@mix.wvu.edu \\
Alexey A. Soluyanov: soluyanov@itp.phys.ethz.ch \\ \\

\vspace*{\fill}

%%% INSERT SI file.... I discovered this way of appending SI in the main_file after spending one full day, and trying so many other things.

%\subfile{SI_arxiv.tex}

\pagestyle{empty}
\afterpage{\null\newpage}
\clearpage  % start a new page

%\includepdf[pages=-, landscape=false, pagecommand={}]{SI_arxiv-compressed.pdf}   
\includepdf[pages={1}, pagecommand={}]{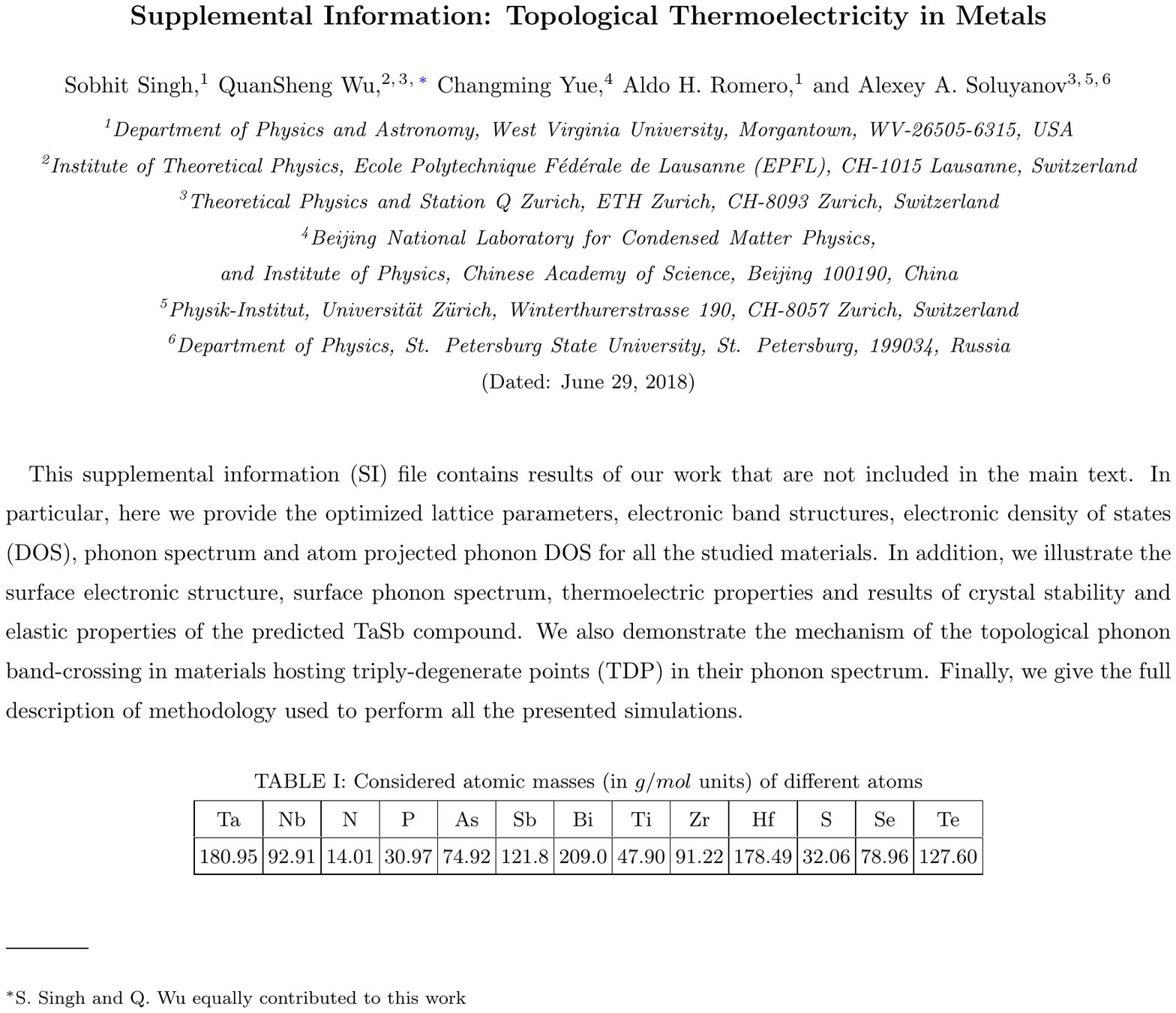}   

\pagestyle{empty}
\afterpage{\null\newpage}
\clearpage  % start a new page
\includepdf[pages={2}, pagecommand={}]{SI_arxiv-compressed.pdf}

\pagestyle{empty}
\afterpage{\null\newpage}
\clearpage  % start a new page
\includepdf[pages={3}, pagecommand={}]{SI_arxiv-compressed.pdf}   

\pagestyle{empty}
\afterpage{\null\newpage}
\clearpage  % start a new page
\includepdf[pages={4}, pagecommand={}]{SI_arxiv-compressed.pdf}   

\pagestyle{empty}
\afterpage{\null\newpage}
\clearpage  % start a new page
\includepdf[pages={5}, pagecommand={}]{SI_arxiv-compressed.pdf}   

\pagestyle{empty}
\afterpage{\null\newpage}
\clearpage  % start a new page
\includepdf[pages={6}, pagecommand={}]{SI_arxiv-compressed.pdf}

\pagestyle{empty}
\afterpage{\null\newpage}
\clearpage  % start a new page
\includepdf[pages={7}, pagecommand={}]{SI_arxiv-compressed.pdf}

\pagestyle{empty}
\afterpage{\null\newpage}
\clearpage  % start a new page
\includepdf[pages={8}, pagecommand={}]{SI_arxiv-compressed.pdf}

\pagestyle{empty}
\afterpage{\null\newpage}
\clearpage  % start a new page
\includepdf[pages={9}, pagecommand={}]{SI_arxiv-compressed.pdf}

\pagestyle{empty}
\afterpage{\null\newpage}
\clearpage  % start a new page
\includepdf[pages={10}, pagecommand={}]{SI_arxiv-compressed.pdf}

\pagestyle{empty}
\afterpage{\null\newpage}
\clearpage  % start a new page
\includepdf[pages={11}, pagecommand={}]{SI_arxiv-compressed.pdf}

\pagestyle{empty}
\afterpage{\null\newpage}
\clearpage  % start a new page
\includepdf[pages={12}, pagecommand={}]{SI_arxiv-compressed.pdf}

\pagestyle{empty}
\afterpage{\null\newpage}
\clearpage  % start a new page
\includepdf[pages={13}, pagecommand={}]{SI_arxiv-compressed.pdf}

\pagestyle{empty}
\afterpage{\null\newpage}
\clearpage  % start a new page
\includepdf[pages={14}, pagecommand={}]{SI_arxiv-compressed.pdf}

\pagestyle{empty}
\afterpage{\null\newpage}
\clearpage  % start a new page
\includepdf[pages={15}, pagecommand={}]{SI_arxiv-compressed.pdf}

\pagestyle{empty}
\afterpage{\null\newpage}
\clearpage  % start a new page
\includepdf[pages={16}, pagecommand={}]{SI_arxiv-compressed.pdf}

\pagestyle{empty}
\afterpage{\null\newpage}
\clearpage  % start a new page
\includepdf[pages={17}, pagecommand={}]{SI_arxiv-compressed.pdf}   

\pagestyle{empty}
\afterpage{\null\newpage}
\clearpage  % start a new page
\includepdf[pages={18}, pagecommand={}]{SI_arxiv-compressed.pdf}   

\pagestyle{empty}
\afterpage{\null\newpage}
\clearpage  % start a new page
\includepdf[pages={19}, pagecommand={}]{SI_arxiv-compressed.pdf}   

\pagestyle{empty}
\afterpage{\null\newpage}
\clearpage  % start a new page
\includepdf[pages={20}, pagecommand={}]{SI_arxiv-compressed.pdf}

\pagestyle{empty}
\afterpage{\null\newpage}
\clearpage  % start a new page
\includepdf[pages={21}, pagecommand={}]{SI_arxiv-compressed.pdf}

\pagestyle{empty}
\afterpage{\null\newpage}
\clearpage  % start a new page
\includepdf[pages={22}, pagecommand={}]{SI_arxiv-compressed.pdf}

\pagestyle{empty}
\afterpage{\null\newpage}
\clearpage  % start a new page
\includepdf[pages={23}, pagecommand={}]{SI_arxiv-compressed.pdf}

\pagestyle{empty}
\afterpage{\null\newpage}
\clearpage  % start a new page
\includepdf[pages={24}, pagecommand={}]{SI_arxiv-compressed.pdf}   

\pagestyle{empty}
\afterpage{\null\newpage}
\clearpage  % start a new page
\includepdf[pages={25}, pagecommand={}]{SI_arxiv-compressed.pdf}

\pagestyle{empty}
\afterpage{\null\newpage}
\clearpage  % start a new page
\includepdf[pages={26}, pagecommand={}]{SI_arxiv-compressed.pdf}

\pagestyle{empty}
\afterpage{\null\newpage}
\clearpage  % start a new page
\includepdf[pages={27}, pagecommand={}]{SI_arxiv-compressed.pdf}

\pagestyle{empty}
\afterpage{\null\newpage}
\clearpage  % start a new page
\includepdf[pages={28}, pagecommand={}]{SI_arxiv-compressed.pdf}

\pagestyle{empty}
\afterpage{\null\newpage}
\clearpage  % start a new page
\includepdf[pages={29}, pagecommand={}]{SI_arxiv-compressed.pdf}

\pagestyle{empty}
\afterpage{\null\newpage}
\clearpage  % start a new page
\includepdf[pages={30}, pagecommand={}]{SI_arxiv-compressed.pdf}

\pagestyle{empty}
\afterpage{\null\newpage}
\clearpage  % start a new page
\includepdf[pages={31}, pagecommand={}]{SI_arxiv-compressed.pdf}

\pagestyle{empty}
\afterpage{\null\newpage}
\clearpage  % start a new page
\includepdf[pages={32}, pagecommand={}]{SI_arxiv-compressed.pdf}

\pagestyle{empty}
\afterpage{\null\newpage}
\clearpage  % start a new page
\includepdf[pages={33}, pagecommand={}]{SI_arxiv-compressed.pdf}

\pagestyle{empty}
\afterpage{\null\newpage}
\clearpage  % start a new page
\includepdf[pages={34}, pagecommand={}]{SI_arxiv-compressed.pdf}

\pagestyle{empty}
\afterpage{\null\newpage}
\clearpage  % start a new page
\includepdf[pages={35}, pagecommand={}]{SI_arxiv-compressed.pdf}

\pagestyle{empty}
\afterpage{\null\newpage}
\clearpage  % start a new page
\includepdf[pages={36}, pagecommand={}]{SI_arxiv-compressed.pdf}

\end{document}